\documentclass[aps,prb,twocolumn,superscriptaddress,showpacs]{revtex4-1} 

\usepackage{amssymb}
\usepackage{graphicx}
\usepackage{amsmath}
\usepackage{dsfont}
\usepackage{bbm}
\begin{document}


\title{Mutation of Andreev into Majorana bound states in long  NS and SNS junctions}

\author{D. Chevallier}
\affiliation{Laboratoire de Physique des Solides, CNRS UMR-8502, Universit\'e Paris Sud, 91405 Orsay Cedex, France}
\author{D. Sticlet}
\affiliation{Laboratoire de Physique des Solides, CNRS UMR-8502, Universit\'e Paris Sud, 91405 Orsay Cedex, France}
\author{P. Simon}
\affiliation{Laboratoire de Physique des Solides, CNRS UMR-8502, Universit\'e Paris Sud, 91405 Orsay Cedex, France}
\author{C. Bena}
\email{cristinabena@gmail.com}
\affiliation{Laboratoire de Physique des Solides, CNRS UMR-8502, Universit\'e Paris Sud, 91405 Orsay Cedex, France}
\affiliation{Institute de Physique Th\'eorique, CEA/Saclay, Orme des Merisiers, 91190 Gif-sur-Yvette Cedex, France}

\date{\today}

\begin{abstract}
We study one-dimensional topological SN and SNS long junctions obtained by placing a topological insulating nanowire in the proximity of either one or two SC finite-size leads. Using the Majorana Polarization order parameter (MP) introduced in Phys. Rev. Lett. 108, 096802 (2012), 
we find that the extended Andreev bound states (ABS) of the normal part of the wire acquire a finite MP:  for a finite-size SN junction the ABS spectrum exhibits a zero-energy extended state which carries a full Majorana fermion, while the ABS of long SNS junctions with phase difference $\pi$ transform into two zero-energy states carrying  {\em two} Majorana fermions with the same MP.
Given their extended character inside the whole normal link, and not only close to an interface, these Majorana-Andreev states can be {\it directly} detected in tunneling spectroscopy experiments.
\end{abstract}

\pacs{
	73.20.-r, 	
	73.63.Nm, 	
	74.45.+c, 	
	74.50.+r, 	
}

\maketitle

\section{Introduction} \label{sec:introduction}

Majorana fermionic states have received a lot of interest \cite{beenakker,alicea} especially because of 
their exotic properties such as non-Abelian statistics that open the perspective of using them for quantum computation.\cite{nayak_dassarma} However, while the observations of some recent experiments are consistent with the existence of a topological phase, \cite{sasaki,kanigel,vednorst,goldhaber} a direct unambiguous detection of the Majorana states is still lacking. A system which is expected to exhibit such states consists of a semiconducting wire with strong spin-orbit coupling  (such as InAs or InSb \cite{spin_orbit_nanowire}) , in the presence of an applied Zeeman field and in the proximity of an s-wave superconductor (SC). \cite{lutchyn_dassarma,oreg_vanoppen} The properties of this system have been extensively studied for an homogeneous SC. \cite{stoudenmire,bena_sticlet,lutchyn_fisher,cool_franz,stanescu_dassarma} However, the Superconductor-Normal (SN) and the Superconductor-Normal-Superconductor (SNS) {\em long} junctions have received less attention. \cite{fu_kane,wimmer,meng_lutchyn,fazio,linder,aguado} Most of the work on SNS junctions has focused on short Josephson junctions dominated by the fractional Josephson effect: the Josephson current has a $4\pi$ periodicity, instead of the usual $2\pi$, and the energy of the Andreev bound states (ABS) goes to zero at $\phi=\pi$. \cite{kitaev,yakovenko,fu_kane08,fu_kane,tanaka09,lutchyn_dassarma,oreg_vanoppen,feigelman,law11,jiang11,meyer,hassler11}

Here we focus on topological SN and SNS {\em long} junctions, in order to study the localization of Majorana states and especially the interplay between the Majorana states and the ABS. While it is rather difficult to obtain analytical solutions for the ABS in this setup, a numerical diagonalization \cite{stoudenmire, stanescu_dassarma,bena_sticlet} of the corresponding  tight-binding Hamiltonian  allows us to 
uncover the main physics of these systems. An important tool in this endeavor is the Majorana polarization, a local topological order parameter introduced in Ref.~\onlinecite{bena_sticlet}.

For a  finite NS junction we find that a localized Majorana state forms, as expected, at the exterior edge of the SC while its Majorana partner does not form at the NS interface, but gives rise to a bound state which is delocalized over the entire non-SC section of the wire. \cite{alicea,fazio,linder,aguado}  
The energy of this state is fixed to zero, independently of external parameters such as the gate voltage, which appears to indicate that this state arises from a modified ABS. 
In order to further investigate this possibility, we focus on a long finite SNS junction where the energy of the ABS can be modified by imposing a phase difference $\phi$ between the two SCs. We find that the two Majorana polarizations (or pseudo-spins) at the two outer ends of the SCs rotate with respect to each other when $\phi$ is modified: they are antiparallel at $\phi=0$ and become parallel at $\phi=\pi$. Since the total Majorana polarization in the system is conserved, the excess of Majorana polarization must be carried by the extended ABS-like states in the normal region, whose energy depends on $\phi$.  When $\phi=\pi$ we indeed find that  a zero energy state is formed, carrying exactly {\em two} Majorana fermions. We call this state a {\it Majorana-Andreev bound state} (MABS).

Based on our analysis we propose a {\it direct} method to detect these MABS using tunneling spectroscopy. Their Majorana character can be tested by their robustness at zero energy with respect to the applied gate voltage, along the lines of, \cite{pillet_joyez,nadya_mason} while their Andreev character can be inferred from the dependence on phase. \cite{pillet_joyez,bena} Such experiments should not be far from feasible (for example using semiconducting wires), in particular for SN junctions (preliminary results have been presented for such  junctions by the Kouwenhoven group \cite{kouwenhoven}). Moreover, they should have more significant results than a measurement of the Josephson current, for which the Majorana effects can be quite subtle. \cite{sasaki,kanigel}

The outline of the paper is as follows. In Sec. \ref{sec:model}, we introduce the model for the long NS and SNS junctions. The LDOS and the Majorana polarization are computed for NS and SNS junctions respectively in Sec. \ref{sec:results_for_NS_junction} and Sec. \ref{sec:results_for_SNS_junction} .  In Sec. \ref{sec:experimental_proposal} we propose an experimental setup to detect directly these MABS. We conclude in Sec. \ref{sec:conclusion}.   

\section{Model} \label{sec:model}

\begin{figure}[h]
	\centering
		\includegraphics[width=6.5cm]{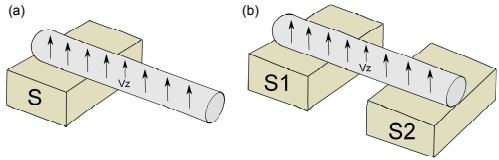}
	\caption{Scheme of the proposed setups : a) a semiconducting nanowire placed partially on the top of a s-wave SC b)  a semiconducting nanowire deposited on  both ends on s-wave SCs.}
	\label{fig:setup}
\end{figure}

Our starting point is a semiconducting nanowire with strong Rashba spin-orbit (SO) coupling.\cite{oreg_vanoppen,stoudenmire}  In the presence of a Zeeman magnetic field and in proximity of a SC, such a wire has been shown to exhibit end Majorana fermionic states. \cite{kitaev,lutchyn_dassarma,oreg_vanoppen}  We focus on using such a wire to make SN or SNS long junctions (See Fig. \ref{fig:setup}).
The superconductivity is induced only in the sections of the wire in contact with the superconductors. To be able to perform a numerical analysis of this system, we model it as a 1D tight-binding chain \cite{stoudenmire}
\begin{align}\label{hamiltonian}
{\cal{H}}&=\sum^N_i \left[(t-\mu)c^\dagger_{i} c_{i}+V_z c^\dagger_{i} \sigma_z c_{i}\right.\notag\\
-&\left.(\frac{t}{2}c^\dagger_{i}c_{i+1}+h.c.)-(i\frac{\alpha}{2}c^\dagger_{i}\sigma_y c_{i+1}+h.c.)\right]\notag\\
+&\sum_{i=1}^{N_1}\Delta_1 e^{i \phi_1} c_{i\uparrow}c_{i\downarrow}+h.c.+\sum_{i=N-N_1+1}^N\Delta_2 e^{i\phi_2} c_{i\uparrow}c_{i\downarrow}+h.c.
\end{align} 
where $N$ is the total number of sites, $\sigma_y$ and $\sigma_z$ are Pauli matrices which act in the spin space, $\mu$ is the chemical potential, $V_z$ the Zeeman field, $\Delta_{1/2}$ is the induced SC pairing and $\alpha$ the SO coupling. We assume a perfect transmission of the junction(s) except in Fig. \ref{fig:Fig1} b) and \ref{fig:Fig4}. 
For the NS junction depicted in Fig. \ref{fig:setup} a), we took $\Delta_2=0$, $N_1=80$ and $N=100$. For the SNS junction depicted in  Fig. \ref{fig:setup} b), we took $\Delta_{1}=\Delta_2=\Delta$, $N=100$ and  $N_1=40$. 
We have checked that our results remain unchanged for much larger lengths of the SC wires (in the regime of parameters we are working with, the spin-orbit length $\xi_{\rm so}$ is much smaller than 
the length of the SCs $(N_1a)$). 
The length of the normal part has been chosen such that a small number (typically a pair) of ABS is formed within the energy range of $\Delta$. We fix the SC phase of the first SC to $\phi_1=0$ and we vary $\phi_2=\phi$. We work in unit of $t=1$, and the lattice constant $a$ is also taken to be equal to $1$.

We focus on two quantities namely the local density of states (LDOS) and the local MP along the x-axis and y-axis which are given by the following expressions~\cite{bena_sticlet}
\begin{align}
\rho_i(\omega)&=\sum^{4N}_{j=1}\sum_{\delta=\uparrow,\downarrow}\delta(\omega-E_j)\left|\alpha^j_{i,\alpha}\right|^2\\
{\cal{P}}^{x}_i(\omega)&=\sum^{4N}_{j=1}\sum_{\delta=\uparrow,\downarrow}\delta(\omega-E_j)2\textrm{Re}(\alpha^{j*}_{i,\delta}\beta^{j}_{i,\delta})\\
{\cal{P}}^{y}_i(\omega)&=\sum^{4N}_{j=1}\sum_{\delta=\uparrow,\downarrow}\delta(\omega-E_j)2\textrm{Im}(\alpha^{j*}_{i,\delta}\beta^{j}_{i,\delta})
\end{align}
where $(\alpha^j_{i\uparrow},\beta^{j}_{i\uparrow},\alpha^{j}_{i\downarrow},\beta^{j}_{i\downarrow})$ are the components of the wave function on site $i$ for the $j^{th}$ eigenstate of the system in the $(c^\dagger_{i\uparrow},c_{i\uparrow},c^\dagger_{i\downarrow},c_{i\downarrow})$ basis. An exact digonalization of the Hamiltonian (\ref{hamiltonian}) allows us to evaluate these two quantities. We focus on the limit when the system is in the topological phase by chosing $V_z=0.4, \Delta=0.3, \mu=0, \textrm{and}\; \alpha=0.2$.
A finite width for the delta functions is introduced in the numerical evaluations responsible
for the finite width of the peaks in the LDOS and MP.
Here we take it to be $0.000015 \hbar v_F/a$ (except in Fig.~\ref{fig:Fig2} where the width is equal to $0.0002 \hbar v_F/a$). 

\section{Results for the NS junction} \label{sec:results_for_NS_junction}

\begin{figure}[ht]
	\centering
		\includegraphics[width=5.5cm]{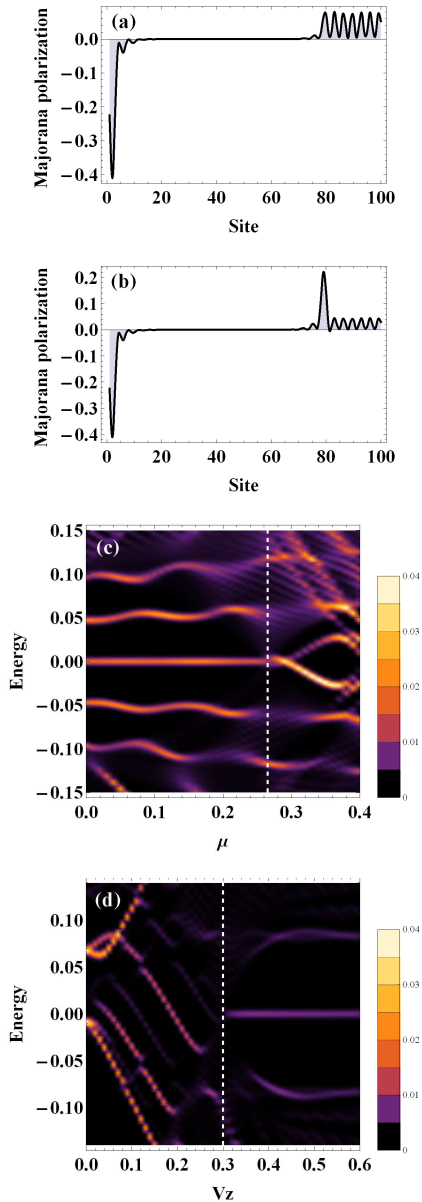}
	\caption{a) The Majorana polarization at zero energy as a function of position. b) The Majorana Polarization at zero energy as a function of position when the coupling between the SC and the normal wire is small $t'=t/2$. c) The LDOS (averaged over a few normal sites) as a function of energy and chemical potential. d) The LDOS (averaged over a few normal sites) as a function of energy and Zeeman field. The white dashed line corresponds to $\tilde{\mu}$ and $\tilde{V}_z$.
}
	\label{fig:Fig1}
\end{figure}

We first remind the reader that in a topological SC wire the spectrum shows two zero-energy Majorana states localized at the two ends of the wire; these two states have opposite MP. The first question that we want to address is how this picture is modified in SN junctions. By diagonalizing the Hamiltonian of this system we find that its spectrum also exhibits two zero-energy modes. In Fig.~\ref{fig:Fig1} a) we plot the zero-energy Majorana polarization  as a function of position. We see that indeed a MP peak is formed at the left end of the topological SC, whose integral is equal to $-1$. \cite{note1} The second zero-energy state does not correspond to a localized mode, \cite{alicea,fazio,linder,aguado} but to a Majorana state uniformly extended over the entire normal section of the wire. \cite{note0} Note that if the normal section of the wire is non-topological, this second Majorana  is localized at the SN interface, on the SC side. \cite{sau_dassarma} Moreover, if the coupling between the SC and the normal wire is small, the Majorana mode also becomes localized at the SN interface (see Fig. \ref{fig:Fig1} b)). We have checked that the integral of the MP for this state is equal to $1$, exactly canceling the MP of the edge mode (the total MP has to be conserved and equal to zero).  

This state is remarkably robust; we have checked that its energy and its MP do not depend on different parameters in the system such as the gate voltage, and the Zeeman field. Thus, in Fig.~\ref{fig:Fig1} c) we plot the LDOS in the normal sector as a function of energy and applied chemical potential. We see that the zero-energy peak is unchanged as long as one remains in the topological phase, but it splits into two peaks when the system exits the topological phase ($\mu>\tilde{\mu}$ where $\tilde{\mu}=\sqrt{V^2_z-\Delta^2}$).\cite{oreg_vanoppen} In Fig. \ref{fig:Fig1} d) we observe a similar behavior for the dependence with the Zeeman field  (here the system exits the topological phase when $V_z<\tilde{V}_z=\sqrt{\Delta^2+\mu^2}$) in agreement with preliminary results presented for a similar setup by the Kouwenhoven group. \cite{kouwenhoven}

\section{Results for the SNS long junction} \label{sec:results_for_SNS_junction}

\begin{figure}[ht]
	\centering
		\includegraphics[width=8.6cm]{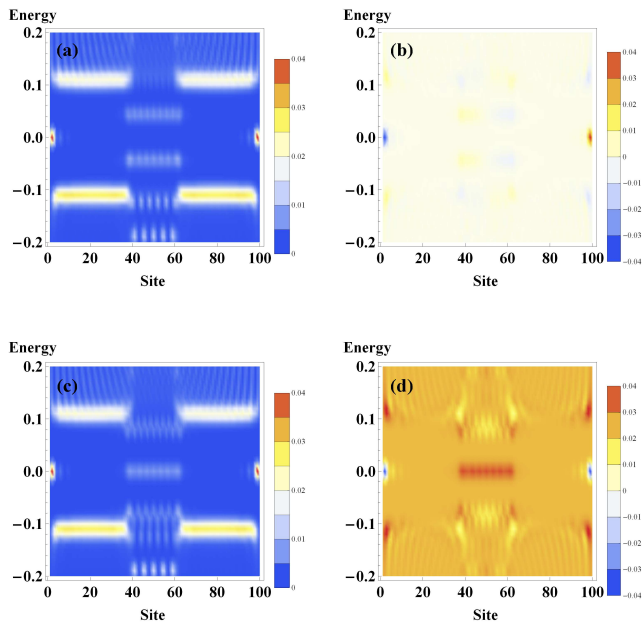}
	\caption{LDOS (a) and x-axis MP (b) for $\phi=0$ and LDOS (c) and x-axis MP (d) for $\phi=\pi$ as a function of energy and position. Note the zero energy states on the exterior ends of the SCs which have equal MP at $\phi=\pi$ and opposite MP at $\phi=0$. Note also the finite MP of the ABS at zero-energy for $\phi=\pi$. Note finally that the effective gap in the SC part is $\hat{\Delta}\approx0.1$ \cite{alicea}.}
	\label{fig:Fig2}
\end{figure}

The nature of this extended Majorana state is intriguing: it seems that an ABS gets pinned to zero in the topological phase, giving thus rise to an extended MABS. In order to clarify the nature of this state, we study a SNS junction  such as the one depicted in \ref{fig:setup} b). For this setup the energy of the ABS can be controlled by modifying the phase difference $\phi$ between the two SCs. The LDOS and the MP along the x-axis are plotted in Fig. \ref{fig:Fig2}  for  $\phi=0$ and $\phi=\pi$ (in these two configurations the MP has no y-component). We can see that for both $\phi=0$ and $\phi=\pi$ two zero-energy states are formed at the two outer edges of the SCs. When $\phi=0$ the MPs of the two end-states are opposite, however, when $\phi=\pi$ the two MPs are oriented in the same direction. The integral of the MP of these states is respectively $-1$ and $1$ for $\phi=0$, and $-1$ for both states at $\phi=\pi$.

As expected, we also observe the formation of ABS in the normal link. In a non-topological SNS junction, the evolution of the ABS with the phase difference follows a standard sinusoidal dependence. \cite{andreev_works, pillet_joyez, bena} For the topological system described here, the energy of the ABS is also affected by the phase difference, in particular we see that at  $\phi=\pi$ the energy of the ABS goes to zero, consistent with previous observations. \cite{sau_dassarma,meyer,meng_lutchyn,fu_kane} Most 
interestingly, we find that the ABS acquire a finite Majorana polarization, which is uniformly distributed along the normal link, and whose integral at $\phi=\pi$ is exactly equal to $2$. \cite{note2} This implies that the zero-energy ABS is a Majorana state: an extended MABS. A similar observation has been made for the states at the interfaces in short junctions. \cite{sau_dassarma, meng_lutchyn, yakovenko}  However, we claim here that in long SNS junctions, localized Majorana states do not form at the interfaces, but instead that the extended ABSs become uniformly Majorana polarized at $\phi=\pi$.

We have checked that at $\phi=\pi$ there are indeed four zero-energy Majorana modes in the spectrum. Moreover, we have verified that the wave function of the zero-energy states has the form  $\alpha(c^{\dagger}_{\uparrow}+c_{\uparrow})$, satisfying the Majorana reality condition. On the other hand, if the SC phase difference is different from zero and $\pi$, the non-zero energy ABS carry a fractional MP and are a mixture of Andreev and Majorana (having non-zero components for all four electron/hole, spin up/down components of the wavefunction) which continuously interpolates between a full Majorana ($\phi=\pi$) and a full Andreev state ($\phi=0$).

\begin{figure}[ht]
	\centering
		\includegraphics[width=4.5cm]{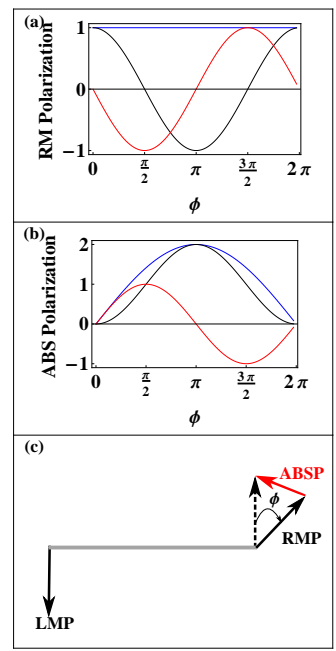}
	\caption{Majorana polarization along the x-axis (black), y-axis (red) and the norm of the MP vector (blue) integrated over the right-end Majorana (a) and over the ABS (b). c) Schematic picture of the Majorana pseudo-spin rotation as a function of the phase difference.}
	\label{fig:Fig3}
\end{figure}

At $\phi=\pi$, when the two Majorana states at the outer edges have parallel pseudospins and a MP of $-1$, the central ABS gets a MP of $2$, in order to keep the total MP of the wire to zero. In Fig.~\ref{fig:Fig3} we plot  the x-axis and y-axis MP as well as the norm of the MP vector as a function of $\phi$ for the right-end mode (integrated over the entire peak), and the Majorana polarization of the ABS (integrated over the entire normal part). The left-end Majorana pseudo-spin does not move since the phase of the left SC does not change. We can clearly see that the MP  vector of the right mode rotates when varying $\phi$ (the phase difference between the SCs is equal in our convention to the angle of rotation $\phi$ described in Fig. \ref{fig:Fig3} c) , thus for $\phi=0$ and $\phi=\pi$ the Majorana polarization along the y-axis is zero). To compensate for this, the central ABS acquires a finite MP, maximal when the phase difference is $\phi=\pi$.  A schematic picture for this phenomenon is presented in Fig.~\ref{fig:Fig3} c).

\begin{figure}[ht]
	\centering
		\includegraphics[width=9cm]{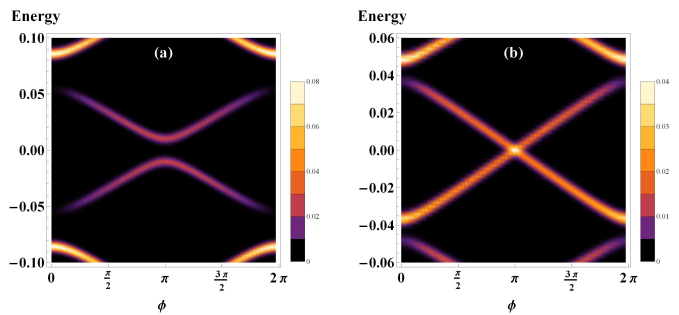}
	\caption{The LDOS in the normal link exhibiting ABS peaks as a function of energy and phase difference in a non-topological (a) and topological (b) state. We take $V_z=0.4$, $\Delta=0.3$, $\alpha=0.2$, and $\mu=0$ in the topological state and $V_z=\alpha=0$, $\Delta=0.3$, and $\mu=1$ in the non-topological state. For clarity we have chosen to open a gap at $\phi=0$ and $\phi=2\pi$ between the two first ABS by changing the SN coupling at the two SN interfaces ($t'_L=0.85$ and $t'_R=1.15$).}
	\label{fig:Fig4}
\end{figure}

We have also performed a detailed check of the dependence of the energy of the ABS on $\phi$. In Fig. \ref{fig:Fig4} we have plotted the LDOS  in the normal section as a function of the energy and the phase difference when the wire is both in a  non-topological (a) and topological phase (b). We can see that, consistent with previous observations, \cite{sau_dassarma,meyer,meng_lutchyn,fu_kane} while in the non-topological phase a gap at $\phi=\pi$ can be easily opened by external parameters such as the gate voltage, etc., in the topological phase the energy of the ABS is always zero at $\phi=\pi$, and the height of the zero-bias peak is doubled with respect to that of the ABS peaks at other values of $\phi$. The crossing of the ABSs at $\phi=\pi$  is consistent with the fractional $4\pi$ periodic Josephson effect previously described for short SNS junctions.\cite{beenakker,alicea,yakovenko}

\section{Experimental consequences} \label{sec:experimental_proposal}
 
Before discussing our experimental proposal, we stress the importance of the extended character of the Majorana states described here. This should make one able to observe them directly spectroscopically, either by STM or by using a weakly coupled normal probe, without the constraint of having to position a real STM tip close to an interface, which may be practically delicate.
Moreover, such an extension of the Majorana state across the whole normal link of a SNS junction makes them easier to manipulate, and eventually use for q-bits: as such, these devices are easily scalable towards a Majorana network, since the junctions do not need to be similar.
Besides, the Majorana pseudo-spin of these states can be easily changed by varying the SC phase, giving us an extra handle to play with such states. 
 
\begin{figure}[ht]
	\centering
		\includegraphics[width=5.5cm]{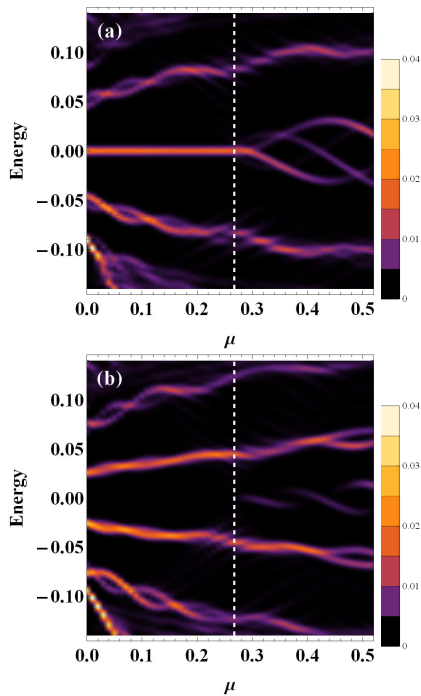}
	\caption{The LDOS as a function of energy and chemical potential for $\phi=\pi$ (a) and $\phi=0$ (b). Note that at $\phi=\pi$ the zero-energy ABS is independent of the gate voltage until the system exits the topological phase $\mu>\tilde{\mu}=\sqrt{V^2_z-\Delta^2}\approx 0.265$ (White dashed line). An extra local gate voltage of $V_g=-0.3$ is introduced here solely on the normal wire to avoid getting close to the bottom of the normal state band structure; the $V_g=0$ case is presented in Fig. \ref{fig:Fig7}. Note also that the transition does not occur exactly at $\tilde{\mu}$ but at a slightly shifted value, due to the applied gate voltage on the N part.
}
\label{fig:Fig6}
\end{figure}

\begin{figure}[ht]
	\centering
		\includegraphics[width=5.5cm]{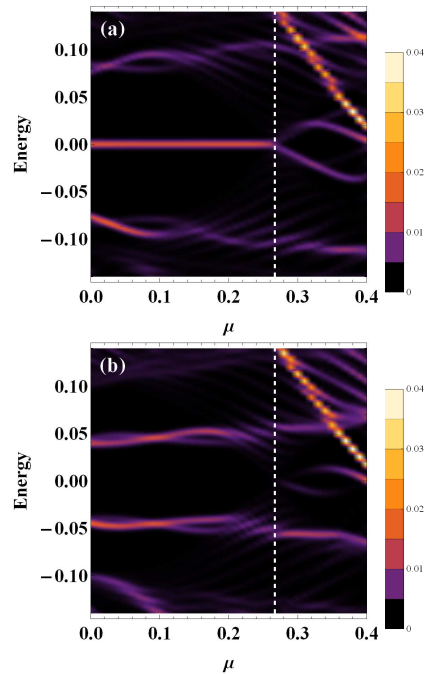}
	\caption{The LDOS in the normal part of the long SNS junction as a function of energy and chemical potential for $\phi=\pi$ (a) and $\phi=0$ (b). Here the chemical potential of the normal section and the SCs is identical which means that the applied gate voltage on the N part is $V_g=0$. The white dashed line corresponds to $\tilde{\mu}$.}
	\label{fig:Fig7}
\end{figure}

We claim that the detection of a zero-energy peak at $\phi=\pi$ in the spectrum of the normal long link would constitute a direct measurement  of the Majorana states, provided two conditions are satisfied: i) the hybridization of the Majorana and the Andreev states into Majorana-Andreev extended states, which can be checked  experimentally by studying  the variation of the ABS with the phase difference between the two SCs; ii) the robustness of the zero-energy state with respect to various parameters in the model such as the gate voltage and the magnetic field.
We believe that one of the easiest and most significant tests would consist of a measure of the LDOS as a function of energy and the chemical potential (gate voltage)  along the lines of Ref. \onlinecite{pillet_joyez}. Notice  that this gate needs to act on both the SC and the normal sections. In Figs. \ref{fig:Fig6} and \ref{fig:Fig7} we present the theoretical prediction for the LDOS in such a measurement. 

Indeed, for $\phi=\pi$, the position and the amplitude of the zero-energy state are not affected by increasing the chemical potential up to a value of $\mu=\tilde{\mu}$ where $\tilde{\mu}=\sqrt{V^2_z-\Delta^2} \approx 0.265$ corresponding to the transition into a non-topological phase. \cite{oreg_vanoppen} When $\mu>\tilde{\mu}$ we recover the well known situation for which the ABS oscillate with the chemical potential. \cite{pillet_joyez,bena}  For much larger chemical potentials the bottom of the band structure becomes apparent (the very bright signature on the top right corner of both plots in Fig. \ref{fig:Fig7}). In order to probe a larger scale of $\mu$ and to observe the classical oscillation of the ABS in the non-topological phase, in Fig. \ref{fig:Fig6} we have considered a non-zero gate voltage $V_g=-0.3$ which acts independently on the normal section of the nanowire.

\section{Conclusion} \label{sec:conclusion}

We have shown how the extended ABS become MABS in long SN and SNS junctions. For SNS junctions we have discovered that the direction of the Majorana pseudo-spin can be controlled by changing the phase difference between the two SCs. The conservation of Majorana polarization implies that the ABSs in the normal link become Majorana-polarized and carry exactly two Majorana fermions with the same MP when the phase difference is equal to $\pi$.
These zero-energy Majorana-Andrev bound states are extremely robust with respect to modifications of various parameters in the system. We propose that a spectroscopic detection of such a zero-energy state along the lines of,  \cite{pillet_joyez} correlated with a measure of its phase dependence and with a test of its robustness with respect to the chemical potential constitutes a direct detection of a Majorana fermionic state.

\acknowledgments

P.S. acknowledges fruitful discussions with R. Aguado. We would like to thank H. Pothier for useful comments on this manuscript. The work of C.B. and D.C. is supported by the ERC Starting Independent Researcher Grant NANO-GRAPHENE 256965.


\begin{thebibliography}{99}
\bibitem{note1} The normalization is different than in \onlinecite{bena_sticlet} where a full Majorana state had a 1/2 polarization.
\bibitem{note0} Note that a similar behavior has been noted for large 2D junction in Ref. \onlinecite{linder}.
\bibitem{note2} Had we considered a ring geometry with a single superconductor, the result would be completely different. We refer the reader to Ref.~\onlinecite{fu_kane}.
\end{thebibliography}
\end{document}